\documentclass[aps,prx,twocolumn,superscriptaddress,footinbib,floatfix]{revtex4-2}
\usepackage{amsmath,amsthm,amssymb}
\usepackage{latexsym}
\usepackage{amscd,graphicx}
\usepackage{graphicx} 
\usepackage{amsmath}
\usepackage{braket}
\usepackage[titletoc]{appendix}
\usepackage{epsfig}
\usepackage{epstopdf}
\usepackage[normalem]{ulem}
\usepackage[dvipsnames]{xcolor}
\usepackage[colorlinks]{hyperref}% http://ctan.org/pkg/hyperref
\hypersetup{
 colorlinks,
 citecolor=blue,
 linkcolor=blue,
 urlcolor=blue}
\usepackage{float}

\definecolor{crimson}{rgb}{0.86, 0.08, 0.24}

\begin{document}

\title{
Quantum Coulomb drag signatures of Majorana bound states
}

\author{Zi-Wei Li}
\affiliation{Department of Physics, Wenzhou University, Zhejiang, 325035, China}

\author {Jiaojiao Chen}
\affiliation{Department of Physics, Wenzhou University, Zhejiang, 325035, China} 

\author{Wei Xiong}
\email{xiongweiphys@wzu.edu.cn}
\affiliation{Department of Physics, Wenzhou University, Zhejiang, 325035, China}
\affiliation{International Quantum Academy, Shenzhen, 518048, China}

\author{Xiao Xue}
\affiliation{International Quantum Academy, Shenzhen, 518048, China}
\affiliation{Hefei National Laboratory, University of Science and Technology of China, Hefei 230088, China}

\author{Zeng-Zhao Li}
\email{lizengzhao@iqasz.cn}
\affiliation{International Quantum Academy, Shenzhen, 518048, China}

\begin{abstract}
{\bf Majorana bound states (MBSs), with their non-Abelian statistics and topological protection, are key candidates for fault-tolerant quantum computation. However, their unambiguous identification in solid-state systems remains a fundamental challenge. Here, we present a theoretical study demonstrating that drag transport in a capacitively coupled double quantum dot system offers a robust and nonlocal probe of weakly coupled MBSs. Using the master equation approach, we investigate both steady-state and transient dynamics and uncover a distinctive signature of MBSs, namely the emergence of pronounced split peaks in the drag transconductance, directly linked to inter-MBS coupling. We further show that the dynamics of quantum coherence is correlated with the emergence and enhancement of MBS-induced split peaks in the drag transconductance. A comparative analysis with trivial subgap states reveals key differences, that is, MBS-induced transconductance peaks are symmetric and exhibit characteristic splitting, while trivial-state features are generally asymmetric and lack such robust splitting behavior. These findings establish experimentally accessible criteria for distinguishing MBSs from trivial subgap states and provide a practical framework for probing Majorana physics through nonlocal transport.}
\end{abstract}

\date{\today}
\maketitle

\vspace{0.3cm}
\noindent
Majorana fermions, quasiparticles with non-Abelian statistics, are regarded as key building blocks for fault-tolerant topological quantum computation\,\cite{Nayak08rmp,Wilczek09nphys,Stern10nature,Leijnse12,Alicea12,Beenakker13annurev,SarmaFreedmanNayak15npjqi,Elliott15rmp,Aguado17rnc,SatoAndo17rpp,AguadoKouwenhoven20,FlensbergOppenStern21,Tanaka24arXiv,Ivanov01prl}. They are predicted to appear as Majorana bound states (MBSs) in topological superconductors, originally in spinless $p$-wave superconductors\,\cite{Kitaev03ap}, and later in systems with spin-orbit coupling, external magnetic fields, and conventional $s$-wave superconductivity\,\cite{LutchynSauSarma10prl,OregRefaelOppen10prl}. Experimentally, one of the most prominent signatures of MBSs is the observation of a zero-bias conductance peak\,\cite{NadjPerge14science,Xu15prl}. However, similar features can also arise from trivial subgap states, including Andreev bound states, posing a major challenge for the unambiguous identification of MBSs\,\cite{PradaKouwenhoven20natrevphys,Finck13prl,LiuSarma17prb,YavilbergGrosfeld19prb,Liu2019,Law2009,Moore2018b,Moore2018,Yu2021,Kells2012}. 

Recently, coupled double quantum dot (cDQD) systems have emerged as promising platforms for probing MBSs\,\cite{Tabatabaei2020, Sierra2019, Keller2016, Kaasbjerg2016, Sanchez2010,Xiao2024}. Such systems allow precise control of charge and spin states and offer flexible tunability via gate voltages and magnetic fields\,\cite{Dvir2023}. Moreover, “poor man’s Majoranas” in cDQD have been demonstrated\,\cite{LeijnseFlensberg12prb, HaafWimmerGoswami24nature}, highlighting their potential as an accessible, tunable,
and scalable platform for studying MBSs' behavior and non-Abelian statistics. 
Besides, the Coulomb drag effect in cDQD, where a current flow through a biased dot induces a measurable drag current in an unbiased dot even in the absence of direct charge transfer, has proven to be a powerful nonlocal probe of MBSs\,\cite{Tabatabaei2020, Sierra2019, Keller2016, Kaasbjerg2016, Sanchez2010}. Recent work has demonstrated that steady-state drag transport can help distinguish Majorana signatures from trivial bound states\,\cite{Xiao2024}.  Nevertheless, more definitive evidence and a deeper understanding of the underlying physics, particularly beyond conventional current measurements and steady-state transport, remain essential. 

To this end, we present a theoretical study of quantum transport in a cDQD system, where a biased dot is capacitively coupled to an unbiased dot connected to weakly hybridized MBSs. Within a master equation framework, we analyze both steady-state and transient dynamics. Our results reveal that, in addition to a pronounced drag current, a characteristic split-peak structure emerges in the drag transconductance, serving as a signature of nonlocal, Majorana-mediated transport. The time-resolved dynamics further elucidate the emergence and stabilization of these signatures, distinguishing them from transient non-Majorana effects. 
Furthermore, we analyze the dynamics of quantum coherence and find that it is correlated with the evolution of the transconductance features, while not directly determining the steady-state signatures. 
Finally, we establish spectroscopic criteria for distinguishing MBSs from trivial subgap states based on peak splitting, symmetry, and robustness within a finite parameter regime. These findings provide a practical framework for identifying MBSs and offer valuable insights into the design of topological quantum devices and non-equilibrium transport in hybrid quantum systems.

\begin{figure*}
	\centering
	\includegraphics[scale=0.68]{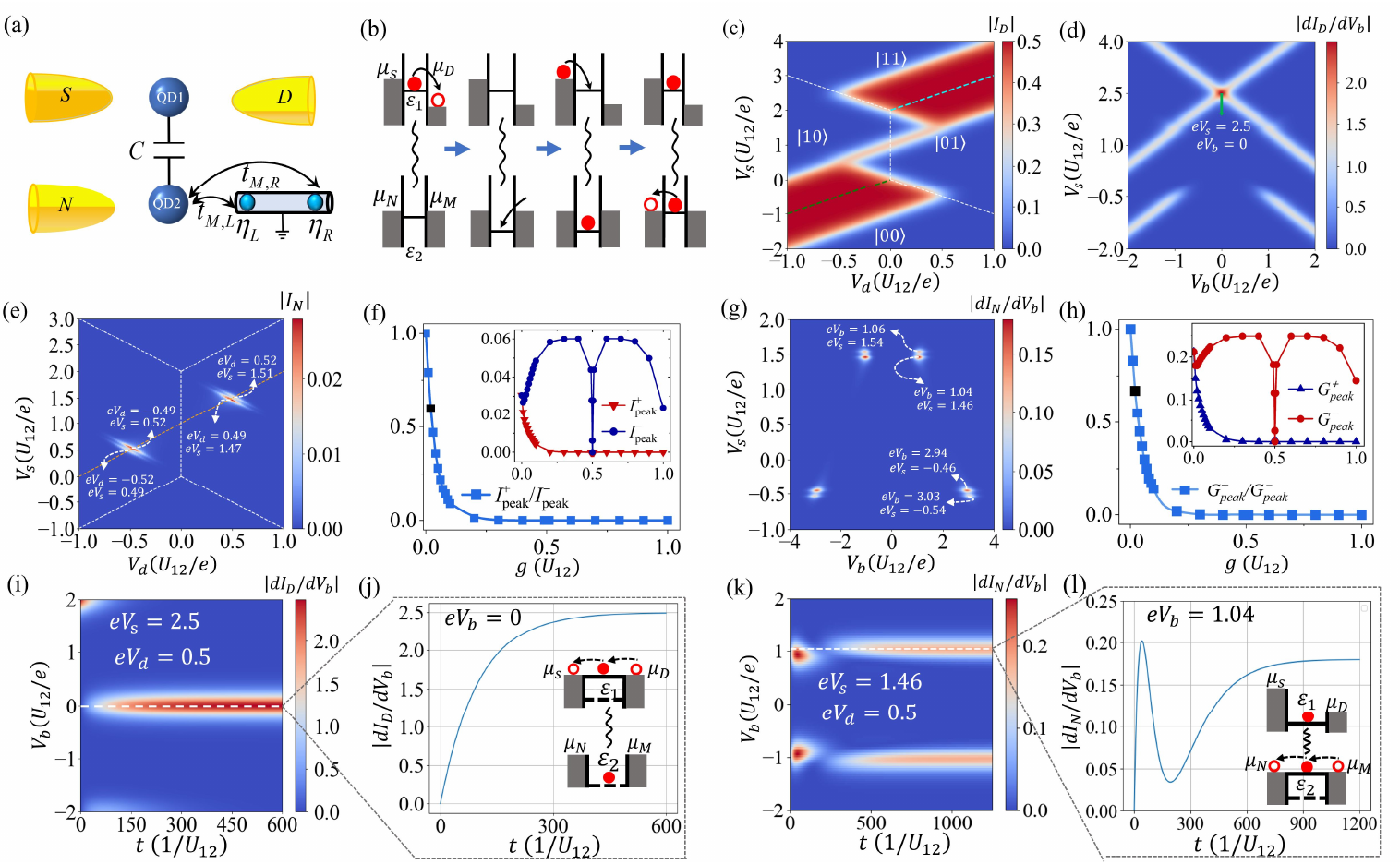} 
	\caption{{\bf Steady-state and transient transport signatures of MBSs.}
    (a) Schematic illustration of the experimental setup for observing MBS-induced drag transport. A bias voltage applied across QD1 generates a drive current from source ($S$) to drain ($D$). Capacitive coupling between QD1 and QD2 induces a drag current through the unbiased QD2, which is connected to a normal metallic lead ($N$) and a proximitized Rashba nanowire hosting two weakly coupled MBSs, $\eta_L$ and $\eta_R$, at its ends. The tunneling amplitudes between QD2 and the MBSs are denoted by $t_{M,L}$ and $t_{M,R}$.
    (b) Illustration of a representative sequence of tunneling events that drive the transition from the state $\left|10\right\rangle$ to $\left|11\right\rangle$, breaking the symmetry between forward and backward transport through QD2.
    (c) Drive current through QD1 as a function of gate voltages $V_s$ and $V_d$ under a fixed bias voltage $V_b = 1.04U_{12}$.
    (d) Differential conductance $dI_D/dV_b$ at fixed $V_d = 0.5U_{12}$, showing a pronounced conductance peak at $(eV_b, eV_s) = (0, 2.5U_{12})$. 
    (e) Drag current through QD2 as a function of $V_s$ and $V_d$ at $V_b = 1.04U_{12}$, revealing two symmetric pairs of MBS-induced current peaks. The upper and lower split peaks in one such pair near $eV_s = 1.5U_{12}$, located at $(eV_d, eV_s) = (0.52, 1.51)U_{12}$ and $(0.52, 1.49)U_{12}$, are labeled as $I^{+}_{\text{peak}}$ and $I^{-}_{\text{peak}}$, respectively. 
    (f) Dependence of the peak current ratio $I^{+}_{\text{peak}} / I^{-}_{\text{peak}}$ on the inter-MBS coupling strength $g$, extracted from data in (e) with additional values of $g$ [see Figs.\,S3(a)–S3(l) in Supplementary Sec.\,II B]. As $g \to 0$, the peak ratio approaches unity, indicating no observable splitting. The rapid suppression of $I^{+}{\text{peak}}$ with increasing $g$ leads to a sharp decline in the ratio. 
    (g) Drag transconductance $|dI_N/dV_b|$ at fixed $eV_d = 0.5U_{12}$ showing four pairs of MBS-induced split peaks centered around $V_s = 1.5U_{12}$ and $V_s = -0.5U_{12}$. In the upper-right pair, the peaks located at $(eV_b, eV_s) = (1.06, 1.54)U_{12}$ and $(1.04, 1.46)U_{12}$ are denoted as $G_{\text{peak}}^+$ and $G_{\text{peak}}^-$, respectively.
    (h) The conductance peak ratio $G_{\text{peak}}^+ / G_{\text{peak}}^-$ as a function of $g$, obtained from data in (g) with extended values of $g$ [see Figs.\,S4(a)–S4(g) in Supplementary Sec.\,II C].
    (i, k) Time evolution of the differential conductance of the drive current $|dI_D/dV_b|$ and the drag transconductance $|dI_N/dV_b|$ (in units of $e^2$), respectively, for $(eV_d, eV_s) = (0.5, 2.5)U_{12}$ and $(0.5, 1.46)U_{12}$.
    (j, l) Voltage slices of (i) and (k) taken at $eV_b = 0$ and $eV_b = 1.04U_{12}$, respectively. The long-time limits of (i) and (k) correspond to the steady-state points in (d) and (g) at $eV_s = 2.5U_{12}$ and $1.46U_{12}$, respectively. 
    $\mu_M$ schematically represents the characteristic energy of the MBSs that serve as an effective bath for QD2. 
    Other parameters are chosen as follows (in units of $U_{12}$): $\mu_S^{(0)} = \mu_D^{(0)} = \mu_N^{(0)} = 0$, $\Gamma_S = \Gamma_D = \Gamma_N = \Gamma_B = \Gamma_A = 0.01$, $k_B T = 0.05$, $g = 0.02$, and $\sigma = 0.01$. Drive and drag currents are given in units of $e\Gamma$, while differential conductances are expressed in units of $e^2$. The cDQD system is initialized in the state $\left| {\psi \left( 0 \right)} \right\rangle  = \left| {00} \right\rangle $.}
    \label{fig1}
\end{figure*}

\vspace{0.6cm}
\noindent\textbf{Results}\\
\noindent\textbf{The model and Hamiltonian}\label{S1}\\
\noindent
We consider a hybrid quantum system consisting of a cDQD with inter-dot capacitive interaction and a superconducting nanowire [Fig.\,\ref{fig1}(a)]. QD1 is driven by a bias voltage applied between two leads (i.e., the source lead $S$ and the drain lead $D$), while QD2 is coupled to a normal lead $N$ and to two spatially separated MBSs located at the ends of a superconducting nanowire with strong Rashba spin-orbit interaction. The Hamiltonian of the proposed system can be written as (setting $\hbar=1$) 
\begin{align}\label{q1}
	H=H_{\rm cDQD} + \sum_\alpha H_\alpha +H_{\rm MBS}+ H_T,
\end{align}
where $H_{\rm cDQD}=\sum_j \varepsilon_j n_j + U_{12} n_1 n_2$ is the Hamiltonian of cDQD, with $\varepsilon_j= - e V_j$ being the energy level of the $j$th QD ($V_j$ is the gate voltage used to control the QD levels), and $U_{\rm 12} = e^2 / 2C$ ($C$ is capacitance) the inter-dot capacitive coupling strength. The operator $n_j = d_j^{\dagger} d_j$ represents the occupation number of the $j$th QD, where $d_j$ denotes the annihilation operator of the $j$th QD. The second term in Eq.\,(\ref{q1}), $H_\alpha=\sum_k (\varepsilon_{k,\alpha} - \mu_\alpha) c_{k,\alpha}^\dag c_{k,\alpha}$ with $\alpha=S,D,N$, denotes the Hamiltonian of the lead $\alpha$. The chemical potentials of the three leads are given by $\mu_S = \mu_S^{(0)} + eV_b/2$, $\mu_D = \mu_D^{(0)} - eV_b/2$, and $\mu_N = \mu_N^{(0)}$, where $e V_b$ is the bias voltage applied to QD1, and $\mu_\alpha^{(0)}$ is the equilibrium chemical potential of lead $\alpha$. $c_{k,\alpha}^\dag$ ($c_{k,\alpha}$) is the creation (annihilation) operator of an electron in the lead $\alpha$. 
The third term, $H_{\rm MBS} = -i g \eta_L \eta_R$, represents the interaction between two separated MBSs with coupling strength $g$, where $\eta_{L(R)}$ denotes the Majorana mode at the left (right) end of the superconducting nanowire. 
The last term, $H_T$, characterizes the tunneling coupling between QD1 and the source and drain leads, and between QD2 and both the normal lead and the two spatially separated Majorana modes. In terms of a fermionic operator defined as $m=(\eta_L - i \eta_R)/\sqrt{2}$, $H_T$ can be explicitly written as
$ H_T =  (\sum_k t_{k,S} c_{k,S}^\dag d_1 +\sum_k  t_{k,D} c_{k,D}^\dag d_1 + \sum_k t_{k,N} c_{k,N}^\dag d_2 + {\rm H.c.}) + t_A (m^\dag  d_2^\dag + m d_2) + t_B (m^\dag d_2 + m d_2^\dag)$, where $t_{k,\alpha}$ is the tunneling amplitude between QD and the lead $\alpha$, $t_A$ is the coupling strength for simultaneous creation (annihilation) of an electron in QD2 and a complex fermion in the nanowire, and $t_B$ is the coupling strength for conventional electron tunneling between QD2 and the fermionic mode. 
Specifically, $t_A = (t_{M,L} - i t_{M,R})/\sqrt{2}$ and $t_B = (t_{M,L} + i t_{M,R})/\sqrt{2}$, where $t_{M,L}$ ($t_{M,R}$) is the tunneling strength between QD2 and the left (right) MBS. 
As the Hamiltonian is expressed with the complex fermion $m$, the original Majorana operators $\eta_{L,R}$ no longer appear explicitly, and the Majorana character of the system is instead reflected in the symmetry and structure of the couplings. 

With the basis spanned by the eigenstates of the isolated DQD, i.e., $\{|a\rangle = |00\rangle, |b\rangle = |10\rangle, |c\rangle = |01\rangle, |d\rangle = |11\rangle\}$, the total system Hamiltonian $H$ in Eq.\,(\ref{q1}) can be rewritten as
\begin{align}\label{q2}
	\mathcal{H} = & \mathcal{H}_{\rm cDQD} + \sum_{\alpha} H_\alpha + H_{\rm MBS}+\mathcal{H}_T,
\end{align}
with 
\begin{align}
	\mathcal{H}_{\rm cDQD}=&\varepsilon_1 |b\rangle\langle b| + \varepsilon_2 |c\rangle\langle c| + (\varepsilon_1 + \varepsilon_2 + U_{12}) |d\rangle\langle d|,\notag\\
	\mathcal{H}_T=&\sum_k \Big[(t_{k,S} c_{k,S}^\dag + t_{k,D} c_{k,D}^\dag)(|a\rangle\langle b| + |c\rangle\langle d|) \notag\\
	& + t_{k,N} c_{k,N}^\dag (|a\rangle\langle c| - |b\rangle\langle d|) \notag\\
	& + (t_A m^\dag + t_B m)(|c\rangle\langle a| - |d\rangle\langle b|) \Big]+ \text{H.c.}, \notag
\end{align}
where $H_\alpha$ is the same as that defined in Eq.\,(\ref{q1}), and $H_{\rm MBS} = g m^\dagger m$. The Hamiltonian in Eq.\,(\ref{q2}) captures the essential features of unidirectional electron transport mediated by quantum tunneling, for example as illustrated in Fig.\,\ref{fig1}(b). Consider a representative transport cycle in which QD1 is initially occupied while QD2 is empty. Due to the inter-dot Coulomb interaction, this occupation electrostatically shifts the energy level of QD2 upward, thereby temporarily suppressing its occupancy. 
Once the electron in QD1 tunnels out to the drain $D$, the energy level of QD2 relaxes, allowing an electron to subsequently tunnel from the lead $M$ into QD2, and then into the normal lead $N$. This process corresponds to the state transition $|b\rangle \rightarrow |d\rangle$, reflecting a directionally correlated sequence of tunneling events mediated by inter-dot Coulomb coupling.
The resulting transport cycle breaks the forward-backward symmetry and induces nonreciprocal transport in the passive dot, emphasizing the intrinsically nonequilibrium nature of the system.

\begin{figure*}
	\centering
	\includegraphics[scale=0.65]{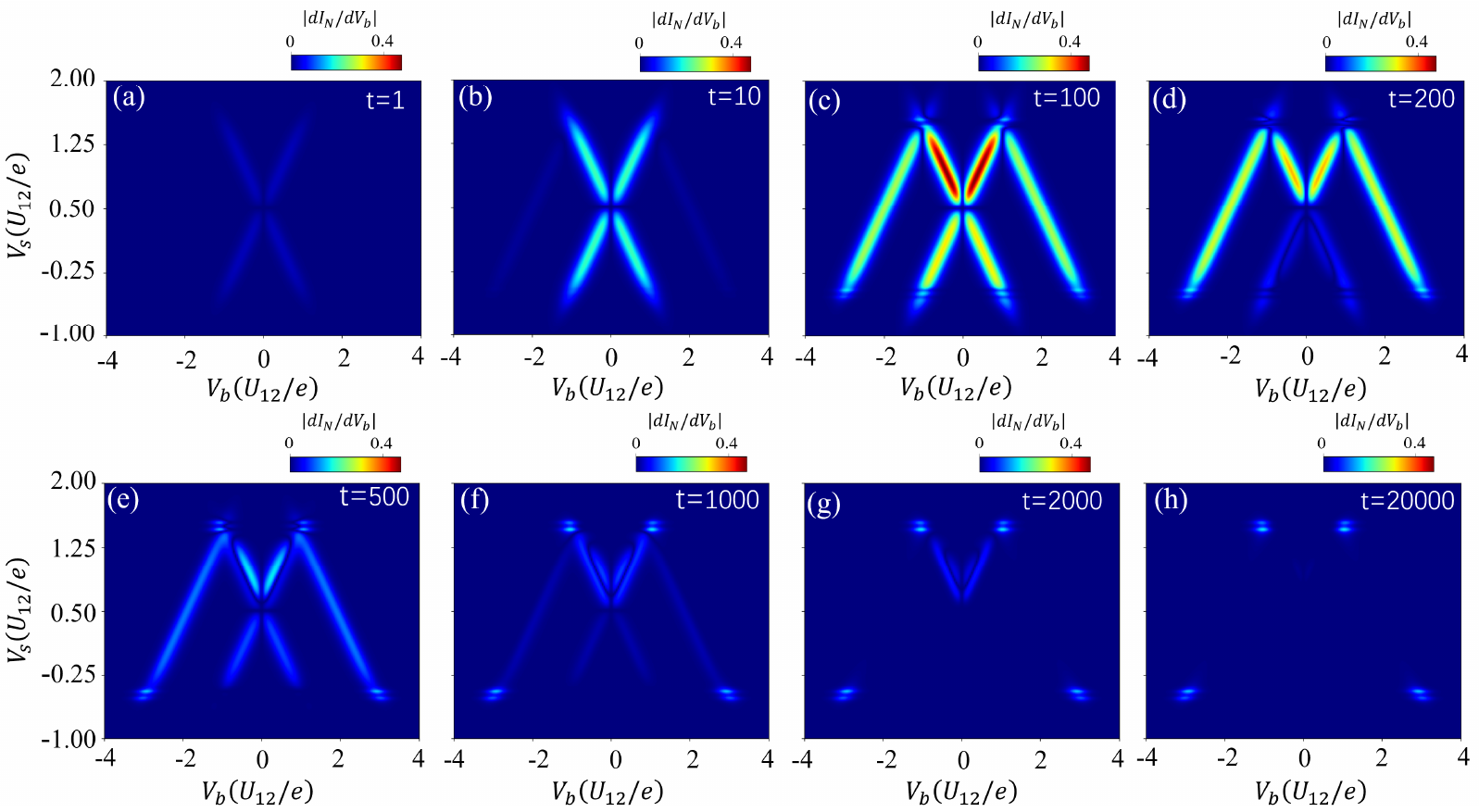} 
	\caption{{\bf The time-evolution characteristics of the drag conductance.}  Panels (a)–(h) depict the temporal evolution of differential drag conductance across the ${V_b}$–${V_s}$ parameter plane ($t$ in units of $1/U_{12}$). Other parameters are the same as those in Fig.\,\ref{fig1} except for $eV_d=0.5 U_{12}$. The initial state for the time evolution is chosen as $\left|\psi\left(0 \right) \right\rangle  = \left|00\right\rangle $. }\label{fig2}
\end{figure*}

\vspace{0.3cm}
\noindent\textbf{Steady-state transport signatures of MBSs}\label{S3}\\
\noindent
{In this work, we use the master equation approach to simulate the dynamics of the proposed system\,({see Supplementary Sec.\,I for details}). We here focus our interest on the steady-state behavior, i.e., $\partial \rho / \partial t = 0$. Solving it with the normalization condition $\sum_i \rho_{ii} = 1$, the occupation probabilities $\rho_{ii}$ in cDQD are directly obtained}. These probabilities enable the evaluation of both the drive and drag currents, along with their corresponding differential conductance. 
Figure\,\ref{fig1}(c) shows the absolute value of the drive current as a function of the gate voltages $V_s = V_2 + V_1$ and $V_d = V_2 - V_1$, under a fixed bias voltage $eV_b = 1.04 U_{12}$. The influence of varying bias voltage on the drag current is analyzed in {Supplementary Sec.\,II A}. 
Along the lines $V_s = V_d$ (green) and $V_s = V_d + 2U_{12}/e$ (cyan), the drive current has a pronounced enhancement, indicating that the tunneling effect is dominant along these two lines. 
The corresponding differential conductance in Fig.\,\ref{fig1}(d) exhibits a prominent zero-bias peak, originating from inter-dot Coulomb interactions that enable electron tunneling from QD1 to the drain lead D even under energy detuning conditions (i.e., $\varepsilon_1/U_{12}=-1$). 

The drag current shown in Fig.\,\ref{fig1}(e) exhibits a distinctive pattern, i.e., two symmetric pairs of split and asymmetric peaks aligned along the red dashed line $V_s = V_d + U_{12}/e$. This feature serves as a characteristic signature of two weakly coupled and spatially separated MBSs, consistent with recent findings reported in Ref.\,\cite{Xiao2024}.
Although the specific parameter choices differ slightly, our results based on the Lindblad master equation are in good agreement with those obtained using the rate equation approach in Ref.\,\cite{Xiao2024}, indicating that quantum coherence plays a limited role in determining the steady-state drag current.
More importantly, we precisely identify the peak positions of the upper-right pair in Fig.\,\ref{fig1}(e), for example at $(eV_d, eV_s) = (0.52, 1.51)U_{12}$ and $(0.49, 1.47)U_{12}$, with corresponding peak currents denoted by $I_{\text{peak}}^+$ and $I_{\text{peak}}^-$, respectively. Their ratio $I_{\text{peak}}^+ / I_{\text{peak}}^-$, extracted from Fig.\,\ref{fig1}(e) with additional values of $g$ [see Figs.\,S3(a)-S3(l) in Supplementary Sec.\,II B], is plotted in Fig.\,\ref{fig1}(f). It exhibits a sharp decrease from near unity to nearly zero as the inter-MBS coupling $g$ increases, primarily due to the suppression of $I_{\text{peak}}^+$. 

The drag transconductance in Fig.\,\ref{fig1}(g) displays four distinct pairs of peaks, with representative peak positions identified at, for example, $(eV_b/U_{12}, eV_s/U_{12}) = (1.04, 1.46)$ and $(1.06, 1.54)$, as well as $(2.94, -0.46)$ and $(3.03, -0.54)$ for $V_b > 0$. The observed peak splittings within each pair arise from finite hybridization between spatially separated MBSs, in contrast to the single peaks located at $(1.04, 1.50)$ and $(2.98, -0.50)$ that correspond to an isolated MBS coupled to QD2 [see Supplementary Sec.\,III A and\,III B for details]. 
As a specific example, the upper and lower peaks in the upper-right pair, denoted $G_{\text{peak}}^+$ and $G_{\text{peak}}^-$, exhibit a strongly asymmetric profile. The corresponding peak conductance ratio $G_{\text{peak}}^+ / G_{\text{peak}}^-$, extracted from Fig.\,\ref{fig1}(g) for various values of $g$ [see Figs.\,S4(a)–S4(g) in Supplementary Sec.\,II C], is plotted in Fig.\,\ref{fig1}(h), showing a sharp departure from $1$ to $0$ with increasing $g$. This behavior reflects the growing asymmetry in the transconductance profile driven by enhanced MBS hybridization. In the Supplementary, the steady-state transport in the case of two decoupled MBSs ($g=0$) and a single isolated MBS are also given for comparsion (see Fig.\,S9 in Sec. III C).

\vspace{0.3cm}
\noindent\textbf{Transient transport signatures of MBSs}\label{S3}\\
\noindent
Next, we examine the transient dynamics to identify the signatures of MBSs beyond the steady state. While the time evolution of the drive current is discussed in {Supplementary Sec.\,II D}, Fig.\,\ref{fig1}(i) shows the evolution of the differential conductance $dI_D/dV_b$ at fixed gate voltages $eV_s = 2.5U_{12}$ and $eV_d = 0.5U_{12}$. A pronounced zero-bias conductance peak gradually develops and saturates around $t = 600$, consistent with the steady-state value observed at the same $V_s$ in Fig.\,\ref{fig1}(d). This steady-state correspondence is further confirmed by the voltage slice at $eV_b = 0$ [indicated by the white dashed line in Fig.\,\ref{fig1}(i)], which is plotted in Fig.\,\ref{fig1}(j).
Figure\,\ref{fig1}(k) presents the time evolution of the drag transconductance $|dI_N/dV_b|$ at $eV_s = 1.46U_{12}$ and $eV_d = 0.5U_{12}$, with the voltage slice at $eV_b = 1.04U_{12}$ shown in Fig.\,\ref{fig1}(l). The drag conductance displays nonmonotonic behavior, initially increasing, then decreasing, followed by a secondary rise before stabilizing at long times. This nonmonotonic dynamics results from the interplay between charge redistribution and capacitive coupling mediated by the MBSs under the bias applied to QD1.
The corresponding transient dynamics of the drive and drag currents for these parameters is provided in Supplementary Sec.\,II E.

\vspace{0.3cm}
\noindent\textbf{Dynamical emergence of MBSs}\\
\noindent
In addition to the pronounced peak splitting observed in the steady-state differential conductance [see Fig.\,\ref{fig1}(g)], a hallmark signature of inter-MBS coupling, we further examine the transient dynamics to elucidate the temporal evolution of MBS-mediated drag transport. Specifically, we analyze the time-dependent behavior of the drag transconductance $|dI_N/dV_b|$ as a function of the normalized bias voltage $V_b/(U_{12}/e)$ and the symmetric gate voltage $V_s/(U_{12}/e)$, while keeping the antisymmetric gate voltage fixed at $V_d = 0.5U_{12}/e$. Figure\,\ref{fig2} presents a sequence of temporal snapshots that trace the system’s evolution from the initial nonequilibrium state to the emergence of a well-defined steady-state transport regime. 

At the initial stage [Fig.\,\ref{fig2}(a)], the transconductance $\left|\mathrm{d}I_N / \mathrm{d}V_b\right|$ exhibits only weak and symmetric features around $V_b = 0$, indicating a minimal transient drag response. As time progresses to $t = 10$ [Fig.\,\ref{fig2}(b)], a distinct X-shaped pattern develops, reflecting the buildup of inter-dot correlations and the onset of nonlocal transport. With further evolution, the transconductance spectrum acquires a characteristic ``M-shaped" profile in the $V_b$–$V_s$ plane [Figs.\,\ref{fig2}(c)–\ref{fig2}(f)]. Pronounced peaks emerge along $eV_s/U_{12} \approx -0.5$ and $1.5$, sharpening over time as resonant energy alignment is achieved between the QD$_2$ levels and the hybridized MBSs. During this intermediate stage ($100 \lesssim t \lesssim 1000$), signatures of Majorana modes become evident, marked by the emergence of nonlocal quantum correlations arising from coherent tunneling processes mediated by inter-dot Coulomb interactions and finite MBS hybridization. At later times [Figs.\,\ref{fig2}(g) and \ref{fig2}(h)], the transconductance amplitude gradually diminishes, signaling relaxation and equilibration toward a stationary regime. By $t = 20000$ (corresponding to $13.164$ ns for $U_{12}=1$ meV \cite{Keller2016,Sierra2019}), the system reaches a steady state characterized by four pairs of asymmetric double peaks in the transconductance spectrum, closely matching the features in Fig.\,\ref{fig1}(g) (see Supplementary Sec.\,II F for the long-time transconductance distribution). 

The emergence of the various structures in Fig.\,\ref{fig2} is governed by multiple competing time scales associated with different physical processes. In particular, the development of the MBS-induced peak structures requires the buildup of coherent correlations mediated by the inter-MBS coupling $g$, which introduces a characteristic time scale on the order of $\sim 1/g$. In the present parameter regime, this time scale is comparable to those set by the dot–lead coupling strengths and thermal broadening. As a result, the MBS-induced features do not appear strictly at asymptotically late times, but instead become clearly resolved only after the initial transient dynamics have relaxed.
In contrast, the X- and M-shaped patterns emerge at earlier stages of the evolution and are primarily governed by dot–lead coupling and thermal effects. The eventual formation of well-defined and stable peak structures at longer times reflects the interplay between coherent Majorana-mediated processes and dissipative relaxation. This separation of dynamical regimes provides further insight into the temporal development of Majorana-induced transport signatures.

The observed time-resolved drag transconductance peaks originate from finite inter-Majorana hybridization, which lifts the degeneracy of the zero modes and signals the emergence of a nonlocal topological steady state. The peak spacing provides a direct dynamical measure of the coupling strength between the two MBSs, serving as a spectral fingerprint of spatially separated Majoranas and the nonlocal correlations they support. For comparison, the transient dynamics of the drag transconductance in the presence of a single isolated MBS are presented in Supplementary Sec.\,III D. Beyond their role as spectral markers, these time-resolved features offer experimentally accessible insight into the real-time evolution of Majorana physics, capturing the interplay between coherent tunneling and dissipative processes that govern drag transport in hybrid QD–MBS systems. These findings open new avenues for time-domain control and detection of Majorana-induced correlations in engineered quantum platforms. 

We note that, while the steady-state behavior is independent of the initial condition, the transient dynamics can depend on the choice of the initial state. In this work, we consider representative initial states, including the empty state $|00\rangle$, which is natural for transport calculations.  We have verified that the main qualitative features shown in Fig.\,\ref{fig2}  remain robust for different initial states, although the early-time evolution may vary.

\begin{figure*}
	\centering
	\includegraphics[scale=0.6]{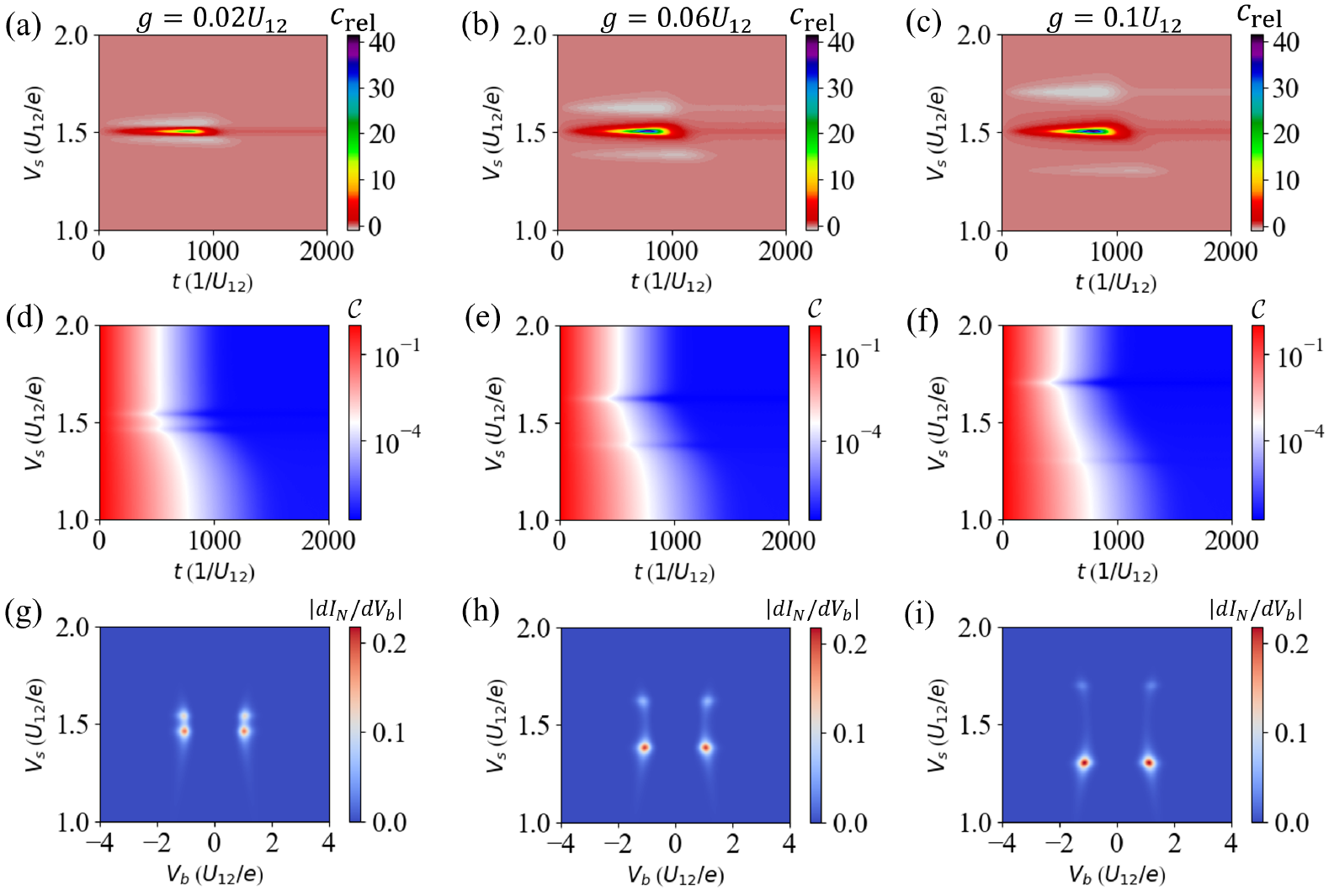} 
	\caption{{\bf Correlation between quantum coherence and conductance characteristics under different inter-MBS coupling strengths.} Panels (a)–(c) show the evolution of the relative quantum coherence $c_{\mathrm{rel}}$ as a function of gate voltage $V_s$ and time $t$, with (a) $g = 0.02U_{12}$, (b) $g = 0.06U_{12}$, and (c) $g = 0.10U_{12}$. Correspondingly, panels (d)–(f) present the evolution of the total off-diagonal coherence $\mathcal{C}$ under the same coupling strengths. Panels (g)–(i) display the steady-state differential conductance $|dI_N/dV_b|$, highlighting its behavior near $V_s \approx 1.5U_{12}/e$. Other parameters are the same as those in Fig.\,\ref{fig1} except for ${V_d} =0.5U_{12}/e$. The initial state for the time evolution is chosen as $\left| {\psi \left( 0 \right)} \right\rangle  = \frac{1}{\sqrt{2}}(|b\rangle + |c\rangle)$.}
    \label{fig3}
\end{figure*}

\vspace{0.3cm}
\noindent\textbf{Quantum coherence}\\
\noindent
Quantum coherence, a defining feature of quantum systems, plays a central role in quantum information processing and transport phenomena\,\cite{Streltsov2017, Baumgratz2014}. It is characterized by the off-diagonal elements of the system’s reduced density matrix, which reflect quantum superpositions. Our earlier findings reveal that, although coherence is present in the system, these off-diagonal terms do not directly contribute to steady-state transport quantities, such as the drive or drag currents and their corresponding conductances\,[see, e.g., Eqs.\,(\ref{eq21}) and (\ref{eq22})]. 
In contrast, coherence plays an important role in the transient regime, where it governs early-time charge dynamics, the emergence of drag currents, and the formation of nonlocal Majorana correlations. Time-resolved observables thus offer a valuable window into coherence effects that remain hidden in the steady state.

To explicitly investigate the dynamics of quantum coherence and their connection to Majorana signatures, we quantify the system’s total coherence using the $l_1$-norm measure\,\cite{Plenio05}, defined as
\begin{eqnarray}
\mathcal{C}(t) = \sum_{i\neq j} |\rho_{ij}(t)|,\,{\text{with}}\,i,j=a,b,c,d,
\end{eqnarray}
where $\rho_{ij}(t)$ are the elements of the reduced density matrix for cDQD, initialized in the coherent superposition state $\left| {\psi \left( 0 \right)} \right\rangle= \frac{1}{\sqrt{2}}(|b\rangle + |c\rangle)$. 
Meanwhile, we specifically examine the coherence characteristics characterized by the off-diagonal element $ \rho_{bc} $, with detailed analysis provided in {Supplementary Sec.\,IV.} 
To assess the impact of inter-MBS coupling, we introduce the relative coherence, 
\begin{eqnarray}
c_{\text{rel}}(g, t) = \mathcal{C}(g,t)/\mathcal{C}(0,t)-1,
\end{eqnarray}
where $\mathcal{C}(0, t)$ serves as the baseline coherence in the limit of negligible Majorana hybridization ($g\rightarrow0$), corresponding to the case of a sufficiently long superconducting nanowire.
Figures\,\ref{fig3}(a)–\ref{fig3}(c) illustrate the evolution of relative quantum coherence $ c_{\mathrm{rel}} $ as a function of time and bias voltage, highlighting the variation of coherence under different Majorana coupling strengths $ g $, with respect to the uncoupled reference value. In Fig.\,\ref{fig3}(a) ($ g = 0.02U_{12} $), a distinct coherence-enhancement region ($ c_{\mathrm{rel}} > 0 $) emerges around $ V_s \approx 1.5U_{12}/e $. As the coupling strength increases to $ g = 0.06U_{12} $ and $ 0.1U_{12} $ [Figs.\,\ref{fig3}(b) and\,\ref{fig3}(c)], this enhanced region becomes more pronounced and expands, indicating that the presence of inter-MBS coupling is associated with enhanced coherence in certain parameter regimes. 
Meanwhile, near the conductance-splitting region centered at $ V_s \approx 1.5U_{12}/e $, a pair of symmetric grey regions appears, corresponding to areas with suppressed coherence ($ c_{\mathrm{rel}} < 0 $). This suggests that, under certain gate-voltage conditions, the coupling to MBS can suppress quantum coherence. 
Notably, as $ g $ increases, the separation between these negatively correlated regions widens, reflecting an expanded range of Majorana-induced influence and a stronger modulation of the system's coherent dynamics.

To further examine this behavior, Figs.\,\ref{fig3}(d)–\ref{fig3}(f) present the evolution of the absolute quantum coherence $\mathcal{C}$, which quantifies the total strength of the off-diagonal elements such as $ \rho_{bc} $ and $ \rho_{cb} $. The coherence rapidly increases at early times and gradually decays due to dissipation caused by the leads. Two narrow features, where coherence sharply decreases to nearly zero, emerge and shift apart as $ g $ increases. This trend is consistent with the variation of $ c_{\mathrm{rel}} $, suggesting that under specific gate-voltage conditions, inter-Majorana coupling can destabilize the off-diagonal elements and suppress coherence.

Since $g$ also affects the transport properties (the full influence of the inter-Majorana coupling strength $g$ on conductance is detailed  in Supplementary Sec.\,II C), Figs.\,\ref{fig3}(g)–\ref{fig3}(i) displays the steady-state differential conductance $\left|\mathrm{d}I_N/\mathrm{d}V_b \right|$ under the same coupling strengths as in Figs.\,\ref{fig3}(a)–\ref{fig3}(c). 
A conductance peak observed around $V_s \approx 1.5U_{12}/e$ [see Fig.\,S3(a) in Supplementary Sec.\,II C]  gradually splits into two symmetric sub-peaks as $g$ increases. This splitting indicates Majorana-induced level hybridization. Notably, the voltage region where this splitting occurs precisely corresponds to where the relative quantum coherence $c_{\text{rel}}$ becomes negative, suggesting that the reduction of coherence occurs in the same parameter regime where Majorana-induced transport features emerge.

These results indicate that the evolution of quantum coherence is correlated with MBS-induced transport features. In particular, regions of enhanced transport (e.g., conductance peaks) coincide with reduced coherence, and vice versa, reflecting a conductance-coherence anti-correlation. This behavior arises from the interplay between coherence, encoded in the off-diagonal elements of the density matrix, and transport processes involving hybridized Majorana states. It therefore reflects a correlation driven by the underlying system dynamics, rather than a direct causal relation between MBS coupling and coherence suppression or enhancement.

\begin{figure}
	\includegraphics[scale=0.38]{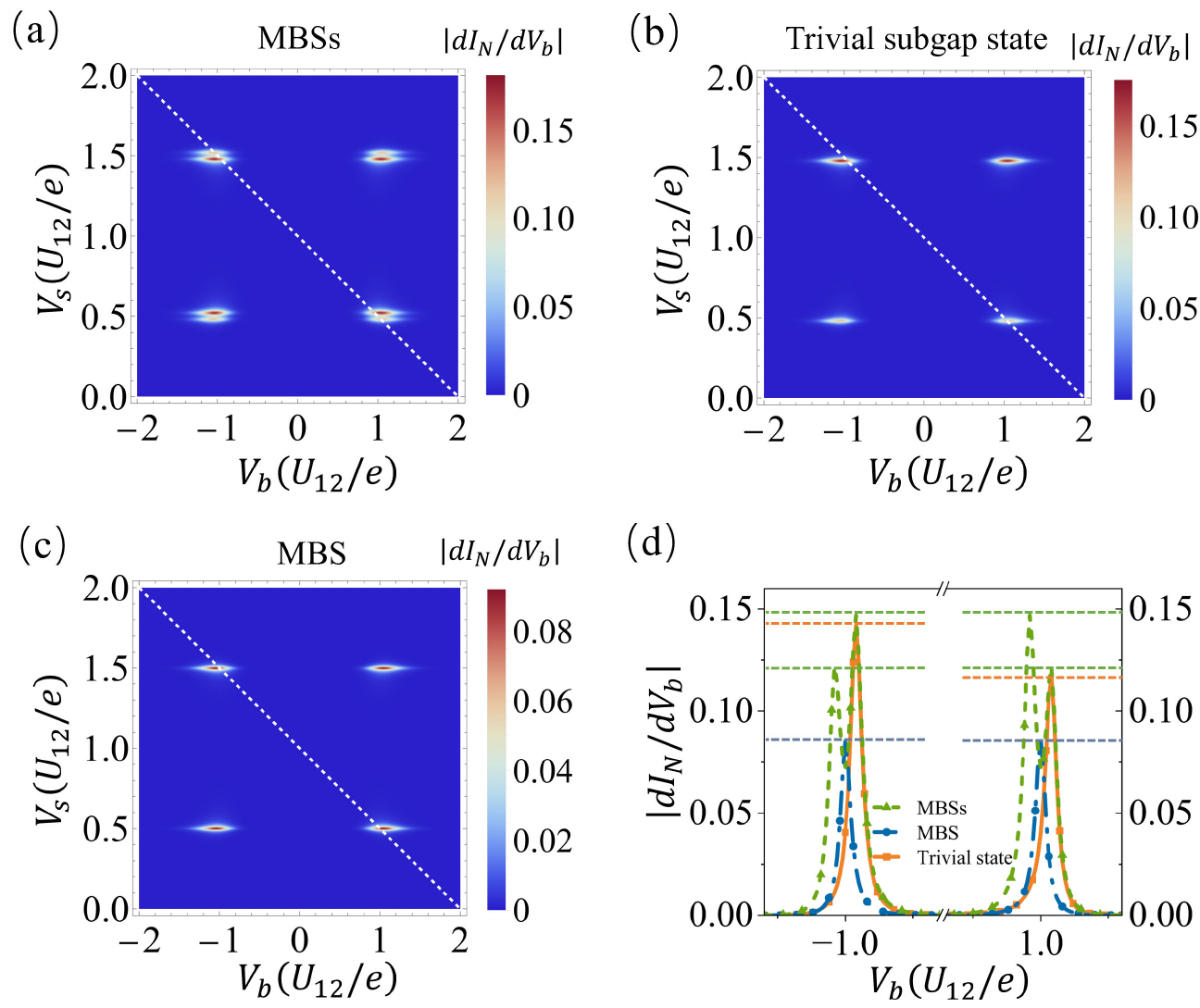}
	\caption{{\bf Distinguishing MBSs from trivial subgap states.} Drag-current differential conductance versus ${V_b}$ and ${V_s}$ for (a) weakly coupled MBSs, (b) a trivial subgap state (modeled as a conventional resonant level), and (c) an isolated MBS, evaluated along $V_s - V_d = 1$ [see the orange dashed line in Fig.\,\ref{fig1}(e)]. (d) Line cuts of the conductance along the white dashed lines in (a)–(c) ($V_s = -V_b/2 + 1$) are compared for different cases: two weakly coupled MBSs (green), an isolated MBS (blue), and a near-zero-energy trivial subgap state (orange). The inter-MBS coupling and the energy of the trivial subgap state are set to $g = 0.02$, and all other parameters are identical to those in Fig.\,\ref{fig1}.}
    \label{fig4}
\end{figure}

\vspace{0.275cm}
\noindent\textbf{Distinguishing Majorana bound states from trivial subgap states}\label{subsec:Andreev}\\
\noindent
A central challenge in tunneling experiments aimed at detecting Majorana bound states (MBSs) lies in distinguishing them from trivial low-energy bound states, which can also produce zero-bias conductance peaks\,\cite{Moore2018,Yu2021}. These trivial states may originate from disorder, quantum confinement, or non-topological subgap excitations, and can mimic certain spectral features of MBSs.  
Within the present framework, we consider a reference case in which the anomalous tunneling term is absent (i.e., $t_A = 0$). In this limit, the system reduces to QD2 coupled to a conventional fermionic level with finite energy and broadening, which serves as a minimal model for a trivial normal bound state (or non-superconducting subgap state). The corresponding tunneling Hamiltonian takes the form
$$
H_T = t_B (m^\dag d_2 + m d_2^\dag),
$$
where the parameter $g$ now represents the energy of the resonant level. 

For direct comparison, Fig.\,\ref{fig4}(a) shows the transconductance spectrum for MBSs along the orange dashed line ($V_s - V_d = 1$) in Fig.\,\ref{fig1}(e), while Figs.\,\ref{fig4}(b) displays the corresponding results for the trivial bound-state case. 
A key distinguishing feature is that, in the trivial case, the drag transconductance peak near $eV_b/U_{12} = \pm 1$ remains unsplit, in stark contrast to the pronounced split-peak structure observed in the MBSs scenario. This absence of splitting reflects the lack of nonlocal hybridization inherent to Majorana pairs. 

We note that the split-peak structure observed for weakly coupled MBSs arises from the hybridization of two nonlocally coupled Majorana modes. Its visibility is governed by both the inter-MBS coupling $g$ and the effective coupling of the dot to the two Majorana components. It is suppressed either when $g$ decreases or when strong coupling asymmetry drives the system toward an effectively local regime. In microscopic nanowire models, such a local regime is often associated with Andreev bound states arising from strongly overlapping or effectively localized Majorana components\,\cite{Ricco19prb}. However, we emphasize that the present model remains within a Majorana-pair framework and does not represent a fully general ABS description.

Further insight is obtained by examining the symmetry of the conductance profiles. In the trivial bound-state case [Fig.\,\ref{fig4}(b) and the orange curve in Fig.\,\ref{fig4}(d)], the peaks exhibit noticeable asymmetry in both height and position, arising from the energy-dependent structure of the resonant level. In contrast, the conductance spectrum associated with an isolated MBS [Fig.\,\ref{fig4}(c) and the blue dash-dotted curve in Fig.\,\ref{fig4}(d)] remains symmetric, reflecting the equal electron–hole weight of Majorana modes. 
For weakly coupled MBSs, shown in Fig.\,\ref{fig4}(a), the transconductance exhibits both symmetry and characteristic peak splitting due to inter-MBS hybridization.

It is important to emphasize that, in contrast to local tunneling spectroscopy, our analysis focuses on drag transconductance in a nonlocal configuration, where the current is measured through the unbiased QD2 while the bias voltage is applied only to QD1. This setup suppresses trivial local contributions and enhances sensitivity to nonlocal transport processes mediated by the coupled Majorana modes. In this configuration, the emergence of symmetric and well-resolved split peaks provides a clear signature of Majorana physics. 
Taken together, these results demonstrate that the combination of peak splitting, symmetry, and robustness of the drag transconductance offers a practical and experimentally accessible set of criteria for distinguishing MBSs from trivial subgap states within this nonlocal transport framework.

\vspace{0.5cm}
\noindent\textbf{Conclusion and discussion}\\
\noindent
In summary, we theoretically investigate quantum transport in a cDQD system within a Markovian master equation framework. By analyzing the drag current and its differential transconductance, we identify characteristic signatures of inter-MBS coupling, in particular the emergence of split peaks that distinguish Majorana-induced transport from that of trivial subgap states within a finite parameter regime. Notably, this splitting develops dynamically over time, providing a time-resolved indicator of Majorana-mediated transport in the sequential tunneling regime. 
We further quantify relative quantum coherence via the $\ell_1$-norm and find that its dynamics is correlated with the evolution of the conductance features, while not directly determining the steady-state transport properties. A comparative analysis between MBS-coupled systems and trivial subgap states modeled as conventional resonant levels reveals clear differences: MBS-induced transconductance peaks exhibit symmetry and characteristic splitting, whereas trivial-state features are generally asymmetric and lack such robust splitting behavior. Together, these results establish a consistent framework for probing Majorana physics and provide guidance for the design of topological quantum devices and the experimental identification of MBSs.

We note that different initial states are employed in this work to probe distinct physical aspects of the system. The empty state $|00\rangle$ is used for transport calculations, while the coherent superposition $(|b\rangle + |c\rangle)/\sqrt{2}$ is adopted to investigate quantum coherence dynamics. Although the transient dynamics may depend on the choice of initial state, the key qualitative features reported in this work remain robust across these representative conditions.

In realistic devices, the couplings between QD2 and the two MBSs, characterized by $t_{M,L}$ and $t_{M,R}$, are generally asymmetric due to their spatial separation along the nanowire. This asymmetry leads to an effective imbalance in the hybridized couplings $\Gamma_A$ and $\Gamma_B$, defined through the combinations $t_{A(B)} = (t_{M,L} \mp i t_{M,R})/\sqrt{2}$, which enter the transport description. To examine its impact, we analyze the case $\Gamma_A \neq \Gamma_B$ in the Supplementary Information (Sec. II G). 
As the asymmetry increases, the split-peak structure is gradually suppressed, reflecting a crossover toward the single-MBS limit where transport is dominated by one Majorana mode. Importantly, the split-peak signature remains clearly observable within a finite and experimentally relevant parameter regime, where both effective couplings are non-negligible and the inter-MBS coupling $g$ exceeds the effective broadening scale (e.g., $\sigma$). This identifies the regime in which the proposed signature can be reliably observed, while strong asymmetry or vanishing $g$ leads to its disappearance.

Our analysis focuses on the sequential tunneling regime, where electron transport occurs through discrete tunneling events between QDs and leads. This approximation is valid under weak QD-lead coupling and thermal energies exceeding tunneling rates, thereby suppressing higher-order processes like cotunneling\,\cite{rudge2018distribution,golovach2004transport,cornean2011cotunneling,PicoCortes2024}. Within this regime, the system dynamics is well captured by a Markovian master equation, which provides a tractable yet accurate framework for describing charge transport and quantum coherence. The Markovian approximation in fact assumes negligible memory effects due to clear timescale separation between system dynamics and environmental correlations. This allows for a simplified treatment of the dissipative evolution and tunneling processes. However, for a strong system-bath coupling or structured environments, non-Markovian effects may become important. In such cases, more general approaches, such as the Nakajima-Zwanzig formalism\,\cite{Nakajima1958, Zwanzig1960, Li14PRA}, could offer improved accuracy by explicitly incorporating memory effects. While our analysis remains within the Markovian regime, the methodology presented here lays a foundation for future extensions to non-Markovian dynamics under stronger coupling conditions.

The results presented in this work are experimentally feasible. Based on existing experiments (e.g., Ref.\,\cite{Keller2016}), differential conductance signals corresponding to current magnitudes on the order of picoamperes (pA) are well within the detection capabilities of current nanowire-quantum dot platforms.
Moreover, the MBS-induced conductance peaks reach approximately $0.15 \, e^2$ (e.g., Fig.\,S6 in Supplementary Sec.\,II F), exceeding the typical peak value of a single MBS, i.e., $\sim0.08 \, e^2$ (Fig.\,S11 in Supplementary Sec.\,III D), and are comparable to experimentally observed signals in similar systems. 
In contrast, trivial subgap states modeled as conventional resonant levels can produce conductance peaks of similar magnitude, but their spectral features differ qualitatively: they generally exhibit asymmetric peak profiles in both height and position and lack the characteristic splitting associated with coupled MBSs. Such differences are readily detectable using standard low-temperature lock-in techniques.
Furthermore, the MBS-induced conductance features are robust against moderate variations in temperature and bias voltage within the relevant parameter regime. The required interaction strengths (such as $U_{12}$) and the inter-MBS coupling $g$ are consistent with values already achieved in InAs/Al nanowire experiments. These considerations indicate that the proposed drag-conductance measurement scheme is well within current experimental reach. 

Our study goes beyond previous work such as Ref.\,\cite{Xiao2024}, which focused exclusively on steady-state drag current, by systematically analyzing both the drag current and its differential transconductance, including their full time-dependent evolution. This broader perspective provides deeper insights into the dynamical behavior of Majorana-mediated transport and coherence in hybrid QD systems.  Moreover, compared to the widely studied zero-bias conductance peak, which is often difficult to interpret due to its sensitivity to local disorder, gating fluctuations, and non-topological subgap states, our approach offers a conceptually distinct and experimentally robust alternative. 
In our setup, the quantum dot directly coupled to the MBSs is kept unbiased, thereby minimizing charge-fluctuation-induced disturbances and eliminating ambiguities associated with direct tunneling. Instead, a finite bias is applied only to a capacitively coupled auxiliary dot, which does not tunnel to the MBSs. This indirect-bias configuration enables nonlocal detection: Majorana signatures appear exclusively through the induced drag current and its transconductance. In particular, the emergence of well-resolved split peaks in the drag transconductance, arising from inter-MBS coupling, serves as a clear and experimentally accessible hallmark of Majorana physics. By avoiding direct perturbation of the MBS channel, this noninvasive scheme enhances robustness against local fluctuations within a finite, experimentally relevant regime and provides a promising pathway toward more conclusive identification of Majorana modes in mesoscopic systems.

It is instructive to relate our results to previous studies on charge transport in quantum-dot-topological-superconductor hybrids. Earlier works have extensively investigated single- and double-dot geometries directly coupled to short topological nanowires, where Majorana signatures are inferred from local or nonlocal conductance under applied bias, governed by dot-Majorana hybridization~\cite{DengMarcus18prb,Prada17prb,Ricco19prb,Majek21prb}. 
In contrast, we consider capacitively coupled quantum dots and probe Majorana physics via Coulomb-drag-induced transconductance, without requiring direct current through the Majorana-coupled subsystem. This noninvasive scheme is sensitive to the nonlocal nature and symmetry properties of the underlying bound states, enabling a clear distinction between Majorana bound states and trivial states, including Andreev bound states and non-superconducting subgap states.  
These approaches are complementary: while prior works rely on direct tunneling spectroscopy, the present mechanism exploits interaction-mediated drag. The latter may reduce back-action and provide an alternative route to detecting Majorana signatures, subject to the realization of sufficiently strong capacitive coupling and sensitive measurements.

\vspace{0.5cm}
\noindent\textbf{Methods}\label{methods}\\
\noindent
We analyze quantum transport in cDQD within the framework of open quantum systems. By employing the master equation approach\,\cite{Gorini1976,Lindblad1976,Breuer2007}, the dynamics of the reduced density matrix $ \rho $ can be governed by (see Supplementary Sec.\,I for details)  
\begin{widetext} 
	\begin{align}
		\frac{{\partial \rho }}{{\partial t}} &= -i[\mathcal{H}_{\rm cDQD}, \rho] \notag\\
		&+\frac{\Gamma_S}{2} \big\{ \bar{f}_S(\varepsilon_1) \mathcal{D}[|a\rangle\langle b|] \rho + f_S(\varepsilon_1) \mathcal{D}[|b\rangle\langle a|] \rho 
		+ \bar{f}_S(\varepsilon_1+U_{12}) \mathcal{D}[|c\rangle\langle d|] \rho + f_S(\varepsilon_1+U_{12}) \mathcal{D}[|d\rangle\langle c|] \rho \big\} \notag\\
		&+\frac{\Gamma_D}{2} \big\{ \bar{f}_D(\varepsilon_1) \mathcal{D}[|a\rangle\langle b|] \rho + f_D(\varepsilon_1) \mathcal{D}[|b\rangle\langle a|] \rho 
		+ \bar{f}_D(\varepsilon_1+U_{12}) \mathcal{D}[|c\rangle\langle d|] \rho + f_D(\varepsilon_1+U_{12}) \mathcal{D}[|d\rangle\langle c|] \rho \big\} \notag\\
		&+ \frac{\Gamma_N}{2} \big\{ \bar{f}_N(\varepsilon_2) \mathcal{D}[|a\rangle\langle c|] \rho + f_N(\varepsilon_2) \mathcal{D}[|c\rangle\langle a|] \rho 
		+ \bar{f}_N(\varepsilon_2+U_{12}) \mathcal{D}[|b\rangle\langle d|] \rho + f_N(\varepsilon_2+U_{12}) \mathcal{D}[|d\rangle\langle b|] \rho \big\} \notag\\
		&+ \frac{\Gamma_B}{2} \big\{ G_M(\varepsilon_2) \big[ \bar{f}_M(\varepsilon_2) \mathcal{D}[|a\rangle\langle c|] \rho + f_M(\varepsilon_2) \mathcal{D}[|c\rangle\langle a|] \rho \big] \notag\\
		&\quad + G_M(\varepsilon_2+U_{12}) \big[ \bar{f}_M(\varepsilon_2+U_{12}) \mathcal{D}[|b\rangle\langle d|] \rho + f_M(\varepsilon_2+U_{12}) \mathcal{D}[|d\rangle\langle b|] \rho \big] \big\} \notag\\
		&+ \frac{\Gamma_A}{2} \big\{ G_M(-\varepsilon_2) \big[ \bar{f}_M(-\varepsilon_2) \mathcal{D}[|c\rangle\langle a|] \rho + f_M(-\varepsilon_2) \mathcal{D}[|a\rangle\langle c|] \rho \big] \notag\\
		&\quad + G_M(-\varepsilon_2-U_{12}) \big[ \bar{f}_M(-\varepsilon_2-U_{12}) \mathcal{D}[|d\rangle\langle b|] \rho + f_M(-\varepsilon_2-U_{12}) \mathcal{D}[|b\rangle\langle d|] \rho \big] \big\}, \label{eq3}
	\end{align}
\end{widetext}
where $\mathcal{D}[\Lambda]\rho = 2\Lambda \rho \Lambda^\dagger - \rho \Lambda^\dagger \Lambda - \Lambda^\dagger \Lambda \rho$ is the the Lindblad superoperator,  $f_\alpha(\omega) = [1+\exp((\omega-\mu_\alpha)/k_BT)]^{-1}$ and 
$f_M(\omega) = [1+\exp(\omega/k_BT)]^{-1}$ are the Fermi-Dirac distributions for lead $\alpha$ and MBSs, respectively. 
The tunneling rate is given by $\Gamma_\alpha = 2\pi G_\alpha |t_\alpha|^2$, 
with $G_\alpha$ denoting the density of states of lead $\alpha$. 
The effective couplings between QD2 and MBSs are 
$\Gamma_{A(B)} = 2\pi |t_{A(B)}|^2$. 
For convenience, we introduce $\bar{f}_{\alpha(M)} = 1 - f_{\alpha(M)}$. In our calculation, we model the density of states of MBSs using a Lorentzian distribution\,\cite{Gibertini2012,Xiao2024},
\begin{align}
G_M(\omega) = \frac{\sigma^2}{(\omega - g)^2 + \sigma^2},	
\end{align}
where $ \sigma $ denotes the broadening parameter and $ g $ is the inter-MBS coupling strength.

To characterize the transport properties of the system, we distinguish between two types of currents: the drive current and the drag current. The drive current refers to the charge current flowing through QD1 into the drain electrode under an applied bias between the source and drain leads. In contrast, the drag current denotes the current flowing through QD2 into its normal lead, which is induced solely by the capacitive coupling to QD1, without applying a direct bias to the Majorana-coupled subsystem.
Using the elements $\rho_{ij}$ of the reduced density matrix $\rho$ in Eq.~(\ref{eq3}), where the diagonal components represent occupation probabilities and the off-diagonal components encode quantum coherence, the drive current flowing into the drain lead through QD1 is given by\,\cite{Harbola2006, Bush2021}
\begin{align}
	I_D &= e\Gamma_D \left[ \rho_{bb} \bar{f}_D(\varepsilon_1) + \rho_{dd} \bar{f}_D(\varepsilon_1+U_{12}) \right] \notag\\
	&\quad - e\Gamma_D \left[ \rho_{aa} f_D(\varepsilon_1) + \rho_{cc} f_D(\varepsilon_1+U_{12}) \right], \label{eq21}
\end{align}
while the drag current flowing into the normal lead through QD2 is
\begin{align}
	I_N &= e\Gamma_N \left[ \rho_{cc} \bar{f}_N(\varepsilon_2) + \rho_{dd} \bar{f}_N({\varepsilon_2+U_{12}}) \right] \notag\\
	&\quad - e\Gamma_N \left[ \rho_{aa} f_N(\varepsilon_2) + \rho_{bb} f_N(\varepsilon_2+U_{12}) \right]. \label{eq22}
\end{align}
This distinction allows us to probe nonlocal transport effects, as the drag current reflects interaction-mediated correlations rather than direct charge injection. 
It is important to note that the direction of the drag current is not universal and depends sensitively on system parameters. In particular, the sign of $I_N$ can change with the quantum-dot energy levels, coupling strengths, and applied bias, and is therefore not necessarily opposite to the drive current. This behavior reflects the competition between interaction-mediated transport channels and the spectral properties of the underlying bound states. Figures S1 and S2 in the Supplementary Information illustrate representative parameter regimes in which the drag current may flow either in the same direction as or opposite to the drive current. 
To characterize transport response, we calculate the differential conductance $ dI_D/dV_b $ and $ dI_N/dV_b $ as functions of the applied bias $ V_b $. These quantities measure how sensitively the drive and drag currents respond to voltage modulation and serve as key diagnostics for identifying the underlying transport mechanisms. 

For numerical calculations, we adopt the inter-dot Coulomb interaction $ U_{12} $ as the energy unit. We consider a symmetric lead configuration with chemical potentials $ \mu_S = \mu_S^{(0)}+ eV_b/2 $, $ \mu_D = \mu_D^{(0)}-eV_b/2 $ and symmetric coupling strengths $ \Gamma_S = \Gamma_D$. The system is assumed to be in the weak-coupling, low-temperature regime where $ \Gamma_\alpha \ll k_B T \ll U_{12} $.

\vspace{0.1cm}
\noindent
\textbf{Funding}\\%\textbf{Acknowledgements}\\
This work was supported by the Natural Science Foundation of Zhejiang Province (GrantNo. LY24A040004), Zhejiang Province Key R\&D Program of China (Grant No. 2025C01028),  and Shenzhen International Quantum Academy (Grant No. SIQA2024KFKT010).\\

\vspace{0.1cm}
\noindent
\textbf{Author contributions}\\
Z.W.L. performed the calculations. Z.Z.L. and W.X. conceived and supervised the project. Z.W.L., J.C., W.X., X.X., and Z.Z.L. contributed to discussions and writing of the manuscript.  \\

\vspace{0.1cm}
\noindent
\textbf{Competing interests} \\
The authors declare no competing interests.\\

\vspace{0.1cm}
\noindent
\textbf{Data availability}\\
\noindent All other data that support the plots within this paper and other findings of this study are available from the corresponding authors upon reasonable request.
%(https://zenodo.org/records/12732420)

\end{document}